\documentclass[journal]{IEEEtran}
\usepackage{amsmath,amsfonts}
\usepackage{algorithmic}
\usepackage{algorithm}
\usepackage{array}
\usepackage[caption=false,font=footnotesize,labelfont=rm,textfont=rm]{subfig}
\usepackage{textcomp}
\usepackage{stfloats}
\usepackage{url}
\usepackage{verbatim}
\usepackage{graphicx}
\usepackage{cite}
\usepackage{booktabs}
\usepackage{multirow}
\usepackage{makecell}
\usepackage{hyperref}
\usepackage{float}

\begin{document}
\bstctlcite{IEEEexample:BSTcontrol}


\title{Over-the-Air Adversarial Attack Detection: \\ from Datasets to Defenses}


\author{Li Wang,~\IEEEmembership{Student Member,~IEEE,} Xiaoyan Lei,~\IEEEmembership{Student Member,~IEEE,} Haorui He, \\ Lei Wang,~\IEEEmembership{Senior Member,~IEEE,} Jie Shi,~\IEEEmembership{Senior Member,~IEEE} and Zhizheng Wu,~\IEEEmembership{Senior Member,~IEEE}
\thanks{Li Wang, Xiaoyan Lei, Haorui He and Zhizheng Wu are with the School of Data Science, Shenzhen Research Institute of Big Data, The Chinese University of Hong Kong, Shenzhen (CUHK-Shenzhen), Shenzhen 518172,  China (e-mail: liwang1@link.cuhk.edu.cn; xyan\_lei@163.com; hehaorui11@gmail.com; wuzhizheng@cuhk.edu.cn).}
\thanks{Lei Wang and Jieshi are with with Huawei International Pte Ltd, 9 North Buona Vista Drive, \#13-01, The Metropolis Tower 1, Singapore 138588. (e-mail: wang.lei12@huawei.com and shi.jie1@huawei.com).}
}




\markboth{Journal of \LaTeX\ Class Files,~Vol.~14, No.~8, August~2021}%
{Shell \MakeLowercase{\textit{et al.}}: A Sample Article Using IEEEtran.cls for IEEE Journals}

\IEEEpubid{0000--0000/00\$00.00~\copyright~2021 IEEE}

\maketitle


\begin{abstract}
Automatic Speaker Verification (ASV) systems can be used for voice-enabled applications for identity verification. However, recent studies have exposed these systems' vulnerabilities to both over-the-line (OTL) and over-the-air (OTA) adversarial attacks. Although various detection methods have been proposed to counter these threats, they have not been thoroughly tested due to the lack of a comprehensive data set.
To address this gap, we developed the AdvSV 2.0 dataset, which contains 628k samples with a total duration of 800 hours. This dataset incorporates classical adversarial attack algorithms, ASV systems, and encompasses both OTL and OTA scenarios. Furthermore, we introduce a novel adversarial attack method based on a Neural Replay Simulator (NRS), which enhances the potency of adversarial OTA attacks, thereby presenting a greater threat to ASV systems. To defend against these attacks, we propose CODA-OCC, a contrastive learning approach within the one-class classification framework. Experimental results show that CODA-OCC achieves an EER of 11.2\% and an AUC of 0.95 on the AdvSV 2.0 dataset, outperforming several state-of-the-art detection methods.

\end{abstract}

\begin{IEEEkeywords}
Adversarial attack, over-the-air, over-the-line, automatic speaker verification
\end{IEEEkeywords}

\section{Introduction}

\IEEEPARstart{A}{utomatic} Speaker Verification (ASV) systems confirm speaker identities by analyzing voice characteristics~\cite{bai2021speaker} in applications like voice assistants, in-vehicle control systems, and phone banking. The accuracy and reliability of ASV systems are crucial for their widespread adoption. 
For example, in phone banking, incorrect identification by an ASV system could lead to serious consequences, such as financial loss or privacy breaches for users. Therefore, the accuracy and robustness of ASV systems is vital to ensuring the secure operation of these real-world applications.

Unfortunately, recent studies have exposed vulnerabilities in ASV systems~\cite{tan2022adversarial,yi2023audio} to adversarial audio attacks. These attacks only involve adding imperceptible perturbations to boni fide audio samples, but can easily deceive ASV systems, leading them to misidentify the speaker~\cite{zhang2020black}.
Even more concerning, our previous research demonstrates that these adversarial audio attacks may retain their threat to ASV systems even after being transmitted over the air~\cite{wang2024advsv,10447811}.

To defend against adversarial attacks, researchers have proposed various detection methods~\cite{DBLP:conf/dada/WuKMML23,tan2022adversarial,9551961,villalba21_interspeech,li20r_interspeech,wu2022adversarial,wu2021improving,10159518}. \textit{However, due to the lack of a comprehensive dataset, most methods have only been tested on separate datasets with limited adversarial attack scenarios, lacking fair comparisons across different approaches.} Therefore, there is an urgent need for a comprehensive dataset that encompasses diverse adversarial attack scenarios to thoroughly evaluate these detection methods.

In response, this paper presents AdvSV 2.0, a comprehensive adversarial attack dataset. AdvSV 2.0 encompasses classical adversarial attack algorithms applied to various ASV systems, including both over-the-line and over-the-air transmitted adversarial samples. It also considers replay devices and mobile recording devices of varying fidelity levels.

Additionally, given that the threat of adversarial attacks may diminish after direct replay, we innovatively incorporates a generative neural network to simulate the over-the-air replay process and generate adversarial attack samples in an end-to-end manner. This novel approach  optimizes perturbations for robustness against transmission distortions, significantly enhancing the effectiveness of adversarial attacks even after OTA transmission.

\IEEEpubidadjcol

The current state-of-the-art adversarial sample detection method uses an adversarial purification module to remove perturbations from audio samples while keeping non-adversarial information~\cite{wu2022adversarial}. Changes in ASV scores between the original and purified audio are used as indicators; significant score changes represent adversarial samples. However, current state-of-the-art (SOTA) detection methods often assume a white-box setting, where the ASV model used for detection is identical to the one targeted by the adversarial attack. This assumption significantly limits their real-world applicability. Experimental results show that while these methods perform reasonably well on in-domain samples, they fail to generalize effectively on out-of-domain data, with the EER dropping significantly by at least 10\%~\cite{villalba21_interspeech}.

To address these critical limitations, especially the restrictive white-box assumption and poor generalization on out-of-domain data, we propose CODA-OCC, a novel contrastive domain-aligned one-class adversarial attack detection method. Our approach leverages the concept of one-class classification, which requires only bona fide samples for training, thereby inherently avoiding overfitting to known adversarial samples and the need for a white-box attack assumption. Furthermore, to enhance its generalization capability across diverse domains and unseen adversarial variations, we design a contrastive learning paradigm within the one-class classification framework, which effectively preserves multi-level information in the audio data.

This paper makes the following contributions:

\begin{itemize}
    \item \textbf{AdvSV 2.0, an open-source dataset for adversarial attacks on speaker verification (ASV) systems}. AdvSV 2.0 is highly comprehensive, including 8 targeted ASV models and 4 attack methods. It also considers over-the-air (OTA) scenarios with 3 playback devices and 3 recording devices. The dataset comprises 800 hours and 628K adversarial samples, providing a robust foundation for evaluating the resilience of ASV models against adversarial attacks.

    \item \textbf{To address the inherent challenge of reduced attack performance after OTA transmission, we introduce a neural replay simulator (NRS)-based OTA adversarial attack method}. Experimental results show that NRS improves the absolute success rate of adversarial attacks by an average of 17.8\%, confirming that even after over-the-air transmission, adversarial attacks can maintain a non-negligible success rate (at least 33.5\% as shown in Table ~\ref{tab:nrs adv ota}), thus posing a substantial threat to ASV systems.
    
    \item \textbf{To defend against these powerful and robust adversarial attacks, we propose the Contrastive Domain-Aligned One-Class Classification (CODA-OCC) method for adversarial sample detection}. This method innovatively incorporates the concept of contrastive learning into one-class classification models, preserving the original semantic information of features from pre-trained audio models, thereby enhancing generalization. Experimental results demonstrate that CODA-OCC reduces the EER by an absolute 26.2\% compared to traditional one-class classification and by 8.6\% compared to the baseline, effectively detecting adversarial samples and facilitating deployment in real-world scenarios.

\end{itemize}

\section{Related Work}

\subsection{Adversarial Attacks on Automatic Speaker Verification}
\label{sec:adv on asv}

\subsubsection{Automatic Speaker Verification (ASV)}
An ASV system determines if two speech utterances are from the same speaker using the following criterion:
\begin{equation}
\hat{y}_\text{spk} = \begin{cases}
1, & \text{if } f(x_e, x_v; \theta_{\text{ASV}}) \geq \tau\\
0, & \text{otherwise}
\end{cases}
\label{equ:ASV}
\end{equation}
In this formula, \(x_e\) denotes the enrollment speech sample and \(x_v\) is the evaluation speech sample. The function \(f(\cdot; \theta_{\text{ASV}})\) is the ASV model with parameters \(\theta_{\text{ASV}}\) that extracts speaker embedding vectors from \(x_e\) and \(x_v\) and computes their similarity score. The threshold \(\tau\) is predetermined by the system. If the similarity score \(f(x_e, x_v; \theta_{\text{ASV}})\) exceeds \(\tau\), the system concludes the two utterances are from the same speaker; otherwise, it concludes they are from different speakers.

\subsubsection{Over-the-line Adversarial Attacks on ASV}
Adversarial attacks are a phenomenon in machine learning where carefully crafted, imperceptible perturbations are added to input data, causing the model to make predictions inconsistent with human expectations~\cite{goodfellow2014explaining}. 

Since attackers typically cannot access the enrollment samples, adversarial attacks on ASV systems are created from the evaluation samples \(x_v\). These attacks can be categorized as targeted or untargeted. 

Targeted attacks are designed to mislead the system into verifying a non-target speaker as a specific target speaker, specifically for samples where the true speaker label is different ($y_{spk}=0$). These attacks are more challenging as they require meticulously crafting imperceptible perturbations to \textit{steer} the samples toward a precise target. In contrast, untargeted attacks merely aim for the ASV output to be incorrect, causing the system to misclassify the sample as any speaker identity other than the true one.

In this work, we focus solely on the more challenging targeted adversarial attacks against ASV systems, while untargeted attacks are not considered.
Mathematically, a targeted adversarial sample $x_v^{Adv}$ is crafted by adding an imperceptible perturbation $\delta$ to a clean evaluation sample $x_v$. The magnitude of the perturbation is constrained by a bound $\epsilon$, ensuring it remains undetectable to human ears. The attack is successful if, after adding $\delta$, the ASV system misclassifies the sample by verifying it as the target speaker when its similarity score exceeds a predefined threshold $\tau$. This process can be formulated as follows:
\begin{equation}
x_v^{Adv}=x_v+\delta
\end{equation}
subject to the following conditions:
\begin{equation}
\label{eq:targeted_attack}
s.t.\begin{cases}
f(x_e,x_v;\theta_{ASV}) & < \tau \\
f(x_e,x_v^{Adv};\theta_{ASV}) & \ge \tau \\
|\delta| & < \epsilon
\end{cases}
\end{equation}
Here, $x_e$ denotes the enrollment speech sample, and $f(\cdot;\theta_{ASV})$ is the ASV model with parameters $\theta_{ASV}$ that computes the similarity score between the enrollment and evaluation samples.


In the domain of adversarial attacks on ASV, various methods have been devised to enhance attack stealth and efficacy. FoolHD~\cite{9413760} utilizes a multi-objective loss function to generate minimally perceptible adversarial samples. FakeBob~\cite{chen2019real} employs a novel threshold and gradient estimation technique for effective black-box attacks. Zuo et al.~\cite{zuo22b_interspeech} improve sample generalization with a speaker-specific ensemble method. Additionally, a spectral transformation framework (STA-MDCT)~\cite{10426806} enhances attack transferability and interpretability by modifying voice sample frequency bands and utilizing class activation maps for visualization.

\subsubsection{Over-the-air Adversarial Attack on ASV}
\label{related:OTA}

In certain ASV systems, attackers cannot directly feed over-the-line audio samples into the system. Instead, they must play the audio sample over the air, and the system receives the analog signal transmitted through the physical space and digitizes it. This attack scenario is known as an over-the-air (OTA) attack or replay attack. In this work, we denote the OTA process as $o(\cdot)$.

Under the OTA attack scenario, the decision function of the ASV system can be represented as:
\begin{equation}
\hat{y}_\text{spk} = \begin{cases}
1, & \text{if } f(x_e, o(x_v); \theta_{\text{ASV}}) \geq \tau\\
0, & \text{otherwise}
\end{cases}
\label{equ:ASV_with_OTA}
\end{equation}

Correspondingly, under the OTA attack scenario, the targeted adversarial attack on ASV systems can be formulated as:
\begin{equation}
\begin{aligned}
x_v^{Adv} &= x_v + \delta \\
\text{s.t. } &\begin{cases}
f(x_e, o(x_v); \theta_{\text{ASV}}) < \tau \\
f(x_e, o(x_v^{Adv}); \theta_{\text{ASV}}) \geq \tau \\
|\delta| < \epsilon
\end{cases}
\end{aligned}
\label{equ:adversarial_attack}
\end{equation}

In the context of over-the-air (OTA) adversarial attacks on ASV systems, several studies have particularly focused on enhancing the effectiveness and stealthiness of attacks in real-world, physical environments.
Early works explored various methods to launch OTA attacks: Xie et al~\cite{xie2020realtime}. introduced a real-time, universal adversarial perturbation by modeling room impulse responses to account for physical propagation effects. 
Yuan et al.~\cite{yuan2018commandersong} embedded commands in songs to stealthily manipulate ASR systems through common media channels.
Following these, researchers developed more sophisticated black-box attacks for commercial platforms, where internal system responses are inaccessible~\cite{zheng2021blackbox, 291118}. For instance, Zheng et al.~\cite{zheng2021blackbox} proposed novel black-box attacks on commercial speech platforms, achieving high success rates. Additionally, QFA2SR~\cite{291118} improves transferability through tailored loss functions and time-frequency manipulations, showing significant effectiveness against commercial APIs and voice assistants in a query-free setting.
More recently, efforts have focused on enhancing attack imperceptibility and robustness against real-world distortions. O'Reilly et al.~\cite{oreilly2022effective} developed a less conspicuous adversarial example using adaptive filtering to simplify the attack process while maintaining effectiveness. UTIO~\cite{10078009} introduced a design for creating imperceptible, universal, and targeted adversarial audio examples that maintain high success rates even in OTA scenarios by incorporating psychoacoustic principles to enhance stealth.

\textbf{Existing methods of generating adversarial samples prior to the over-the-air (OTA) process are susceptible to several inherent flaws and challenges.} On one hand, the adversarial perturbation is constrained to have a small magnitude by the definition of adversarial attacks, rendering the generated adversarial samples vulnerable to various factors during the OTA process, thereby diminishing the attack effectiveness. Specifically, environmental noise interference, reverberation effects from different physical environments, and channel attenuation during air transmission can alter or compromise the integrity of the adversarial perturbation. On the other hand, since the adversarial perturbation is crafted before the OTA process, it cannot effectively account for and simulate other unknown microscopic effects that the physical world may impose, further degrading the attack performance. In this work, \textbf{we propose the Neural Replay Simulator Based Over-the-air Adversarial Attacks method} (Section~\ref{sec:NRS based Adv}), which aims to enhance the performance of OTA adversarial attacks.

\subsection{Adversarial Sample Detection on ASV}

One common method for adversarial sample detection involves using simple binary classification models, such as a VGG-like detector~\cite{li20r_interspeech}, to differentiate between adversarial and non-adversarial samples. This approach has shown effectiveness, even when faced with unseen attack settings, though it struggles with robustness against new perturbation methods.Another approach leverages representation learning to classify attacks based on the attack algorithm, threat model, or signal-to-noise ratio, achieving high accuracy but facing challenges in generalizing to unknown attacks~\cite{villalba21_interspeech}.

\begin{figure}[ht]
    \centering
    \includegraphics[width=1\linewidth]{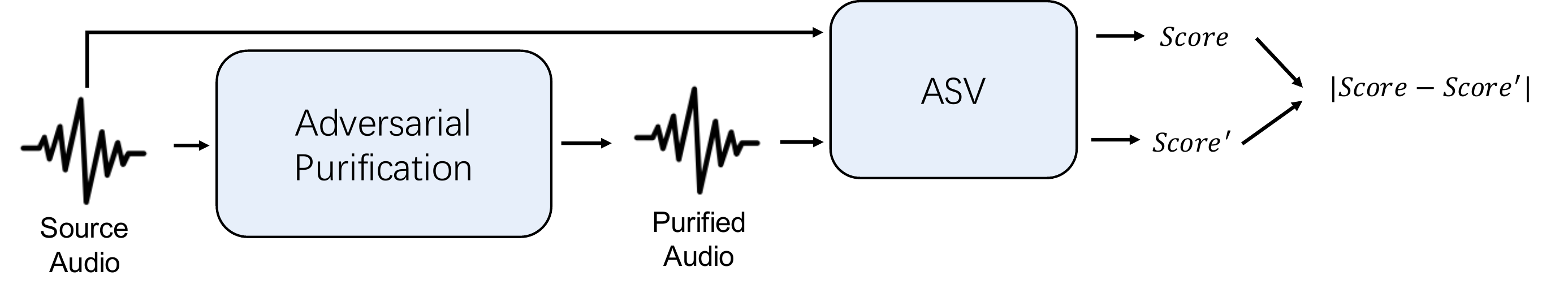}
    \caption{Current mainstream adversarial detection framework. The adversarial purification module removes adversarial perturbations from the source audio, resulting in purified audio. The ASV system computes scores for both the original and purified audio. Significant differences between the scores indicate the presence of adversarial perturbations.~\cite{wu2022adversarial,chen2024neural,wu2021improving}}
    \label{fig:adv_det}
\end{figure}

As show in Fig.~\ref{fig:adv_det}, current mainstream adversarial detection method introduces an adversarial purification module, which removes adversarial perturbations from audio samples while preserving non-adversarial information as much as possible. By observing changes in ASV scores, significant changes indicate the presence of adversarial perturbations. The core idea of this method is to utilize the significant impact of adversarial perturbations on ASV scores, detecting and identifying adversarial attacks through score changes.

Researchers have experimented with vocoders~\cite{wu2022adversarial}, self-supervised models~\cite{wu2021improving}, and codec models~\cite{chen2024neural} as adversarial purification modules. They have also considered integrating multiple adversarial purification models for comprehensive assessment. \cite{10159518} proposed learning a trainable mask for adversarial purification, which retains only information relevant to ASV. The effectiveness of these methods lies in their exclusive training on bona fide samples, thereby failing to reconstruct adversarial perturbations effectively and rendering adversarial attacks ineffective.

However, these methods face several issues. \textbf{They rely on variations in ASV scores, but ASV systems have inherent limitations that can compromise the accuracy and reliability of adversarial sample detection}. Additionally, \textbf{they often assume white-box settings where the attack targets the same model being defended, whereas in reality, adversarial attacks are usually performed in black-box settings.} This discrepancy can result in significantly different ASV score changes, rendering the detection algorithms ineffective.

In this paper, we propose \textbf{Contrastive Domain-Aligned One-Class Classification} (Section~\ref{sec:adv_det}) for adversarial attack detection, which can directly identify adversarial samples without relying on variations in ASV scores and does not assume a white-box setting.

\section{AdvSV 2.0: An Advanced Over-the-air Adversarial Attack Dataset for Speaker Verification}
\label{sec:attack methods}

\subsection{Over-the-line Adversarial Attack Methods}
\subsubsection{Projected Gradient Descent}
Projected Gradient Descent (PGD)~\cite{madry2018towards} is an iterative algorithm for generating adversarial samples. The core idea is to iteratively update the adversarial perturbation of the input sample along the gradient direction of the adversarial objective function, such that the final adversarial sample can cause misclassification by the ASV system. Simultaneously, PGD employs a clipping operation to constrain the magnitude of the adversarial perturbation, ensuring that the generated adversarial samples remain highly imperceptible.

\begin{algorithm}[ht]
\caption{Projected Gradient Descent (PGD) Attack for Targeted ASV Adversarial Samples}
\label{alg:pgd}
\begin{algorithmic}[1]
\REQUIRE Enrollment speech \(x_e\), evaluation speech \(x_v\) from different speakers, step size \(\alpha\), max steps \(S\), \(\epsilon\) for \(L_\infty\) norm ball, ASV model \(f(\cdot, \theta_{\text{ASV}})\) with threshold \(\tau\)
\STATE Initialize \(x_1^{Adv} \gets x_v\) \COMMENT{Initialize adversarial samples}
\FOR{$s=1$ \TO $S$}
\STATE $g \gets \nabla_{x_s^{Adv}} J(f(x_e, x_s^{Adv}; \theta_{\text{ASV}}))$ \COMMENT{Compute gradient}
\STATE $x_{s+1}^{Adv} \gets \text{clip}_{x_v,\epsilon}(x_s^{Adv} + \alpha \cdot \text{sign}(g))$ \COMMENT{PGD update}
\STATE $x_v^{Adv} \gets x_{s+1}^{Adv}$ \COMMENT{Update final adversarial sample}
\IF{$f(x_e, x_{s+1}^{Adv}; \theta_{\text{ASV}}) \geq \tau$}
\RETURN $x_v^{Adv}$ \COMMENT{Early stop if target speaker spoofed}
\ENDIF
\ENDFOR
\RETURN $x_v^{Adv}$ \COMMENT{Return final adversarial sample}
\end{algorithmic}
\end{algorithm}

The iterative process of the PGD algorithm is described in Algorithm~\ref{alg:pgd}. Given an initial non-target speaker's utterance \(x_v\), the algorithm iteratively updates the adversarial sample \(x_s^{Adv}\) by adding a perturbation along the gradient direction of the loss function \(J\) with respect to \(x_s^{Adv}\). The perturbation is scaled by a step size \(\alpha\) and clipped within an \(L_\infty\) norm ball centered at \(x_v\) to ensure imperceptibility. The clipping operation is defined as:

\begin{equation}
\text{clip}_{x,\epsilon}(x') = \min(1, \max(-1, x + \epsilon, \max(x - \epsilon, x')))
\end{equation}
where the perturbation stays below \(\epsilon\) after each iteration.

The algorithm terminates early if the updated adversarial sample \(x_{s+1}^{Adv}\) successfully fools the ASV model into classifying it as the target speaker identity. Otherwise, the final \(x_S^{Adv}\) is returned as the adversarial sample.

\subsubsection{Ensemble PGD}

Recognizing the potential lack of transferability when adversarial samples are crafted to attack a single ASV model, the ensemble PGD algorithm has been explored to generate adversarial samples that can bypass multiple victim ASV models simultaneously. The key idea is to iteratively optimize the adversarial perturbation with respect to an ensemble of victim models until the generated adversarial sample can successfully spoof all of them. The ensemble PGD attack is outlined in Algorithm~\ref{alg:ensemble_pgd}.

\begin{algorithm}[ht]
\caption{Ensemble PGD Attack}
\label{alg:ensemble_pgd}
\begin{algorithmic}[1]
\REQUIRE Enrollment speech \(x_e\), evaluation speech \(x_v\) from different speakers, step size \(\alpha\), max steps \(S\), \(\epsilon\) for \(L_\infty\) norm ball, ensemble of ASV models \(\{f(\cdot, \theta_{\text{ASV}}^1), f(\cdot, \theta_{\text{ASV}}^2), \ldots, f(\cdot, \theta_{\text{ASV}}^K)\}\) with thresholds \(\{\tau_1, \tau_2, \ldots, \tau_K\}\)
\STATE Initialize \(x_{1,0}^{Adv} \gets x_v\) \COMMENT{Initialize adversarial samples}
\FOR{$s=1$ \TO $S$}
\FOR{$k=1$ \TO $K$}
\STATE $x_{s,k}^{Adv} \gets x_{s,k-1}^{Adv}$ \COMMENT{Use previous model's output}
\STATE $g_k \gets \nabla_{x_{s,k}^{Adv}} J(f(x_e, x_{s,k}^{Adv}; \theta_{\text{ASV}}^k))$ \COMMENT{Compute gradient for model $k$}
\STATE $x_{s+1,k}^{Adv} \gets \text{clip}_{x_v,\epsilon}(x_{s,k}^{Adv} + \alpha \cdot \text{sign}(g_k))$ \COMMENT{PGD update for model $k$}
\IF{$f(x_e, x_{s+1,k}^{Adv}; \theta_{\text{ASV}}^k) \geq \tau_k$}
\STATE \textbf{break} \COMMENT{Exit inner loop if model $k$ spoofed}
\ENDIF
\ENDFOR
\STATE $x_{s+1}^{Adv} \gets x_{s+1,K}^{Adv}$ \COMMENT{Use final model's output as current step $x^{Adv}$}
\STATE $x_v^{Adv} \gets x_{s+1}^{Adv}$ \COMMENT{Update final adversarial sample}
\IF{$\forall k, f(x_e, x_{s+1}^{Adv}; \theta_{\text{ASV}}^k) \geq \tau_k$}
\RETURN $x_v^{Adv}$ \COMMENT{Early stop if all models spoofed}
\ENDIF
\ENDFOR
\RETURN $x_v^{Adv}$ \COMMENT{Return final adversarial sample}
\end{algorithmic}
\end{algorithm}

\subsection{Over-the-air Adversarial Attack Methods}
\subsubsection{Direct Over-the-air Adversarial Attack}

As stated in Section~\ref{related:OTA}, in certain ASV systems, attackers cannot directly feed over-the-line audio samples into the system. Instead, they must play the audio sample over the air, and the system receives and digitizes the analog signal. This scenario is known as an over-the-air (OTA) attack or replay attack, denoted as \(o(\cdot)\).

The objective of a targeted OTA adversarial attack is to generate an adversarial sample \(x_v^{Adv}\) such that:
\begin{equation}
\begin{aligned}
f(x_e, o(x_v); \theta_{\text{ASV}}) &< \tau \\
f(x_e, o(x_v^{Adv}); \theta_{\text{ASV}}) &\geq \tau
\end{aligned}
\end{equation}

That is, the original speech sample \(x_v\) is correctly classified, while the adversarial sample \(x_v^{Adv}\) is misclassified as the target speaker after the OTA process.

Factors such as air propagation and microphone recording must be considered to enhance the transferability of the adversarial samples. To address these challenges and improve the performance of OTA attacks, we propose the \textbf{Neural Replay Simulator Based Over-the-air Adversarial Attacks method.} This approach aims to mitigate the impact of the OTA process on adversarial samples, ensuring more reliable and effective attacks against ASV systems.

\subsubsection{Neural Replay Simulator Based Over-the-air Adversarial Attack}
\label{sec:NRS based Adv}
\begin{figure}[ht]
    \centering
    \includegraphics[width=1\linewidth]{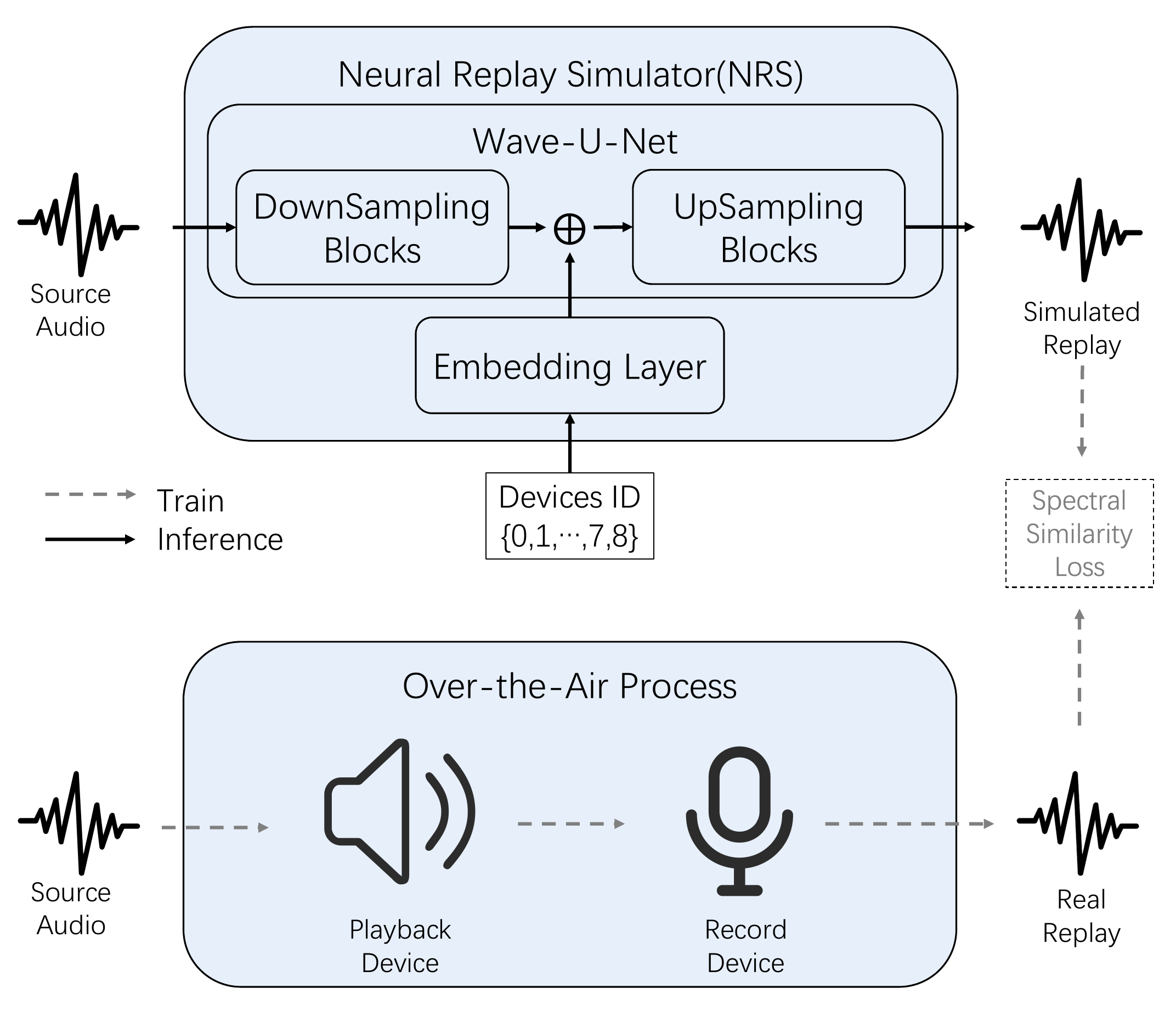}
    \caption{Architecture of the Neural Replay Simulator (NRS).}
    \label{fig:unet}
\end{figure}

As noted in the previous section, existing over-the-air (OTA) adversarial attack methods fail to adequately account for the substantial impact of the subsequent OTA process on the adversarial perturbations during the generation of adversarial samples, as mentioned in Section~\ref{related:OTA}. To tackle this issue, we propose simulating the OTA replay process using a Neural Replay Simulator (NRS). Specifically, we define this simulation as a speech generation task~\cite{amphion,emilia}, where the replay information generated by the NRS is integrated into the adversarial optimization process. This approach enables the generated adversarial samples to endure the distortions introduced by the OTA process.

As illustrated in Fig.~\ref{fig:unet}, we employ the Wave-U-Net~\cite{stoller2018wave} architecture to construct the Neural Replay Simulator (NRS). This model is specifically designed to take original audio recordings as input and generate their anticipated replayed versions post the over-the-air (OTA) transmission process. A unique feature of the NRS is its ability to simulate specific playback and recording device combinations through the use of a Replay ID, allowing precise control over the modeling of different OTA scenarios.
We employed the Multi-Scale Spectral Loss (MSSL)\footnote{\href{https://github.com/babysor/MockingBird}{https://github.com/babysor/MockingBird}}, which is an L1 loss computed on the multi-resolution short-time Fourier transform (STFT) of the input and target signals. The MSSL is defined as:
\begin{equation}
\begin{split}
\mathcal{L}_\text{MSSL} = \frac{1}{N}\sum_{i=1}^N \big( \mathcal{L}_\text{SC}^{(i)} + \mathcal{L}_\text{Mag}^{(i)} \big)
\end{split}
\end{equation}
where $N$ is the number of STFT resolutions, $\mathcal{L}_\text{SC}^{(i)}$ is the spectral convergence loss, and $\mathcal{L}_\text{Mag}^{(i)}$ is the log STFT magnitude loss for the $i$-th STFT resolution.
During its development, the NRS undergoes extensive pre-training on a substantial dataset composed of parallel data, which includes pairs of clean audio recordings and their corresponding versions that have been replayed through various OTA conditions. This comprehensive pre-training enables the NRS to accurately predict the effects of OTA transmission on audio quality and integrity, ensuring that the simulator can effectively recreate the diverse range of acoustic environments encountered in real-world applications.

\begin{figure*}
    \centering
    \includegraphics[width=1\linewidth]{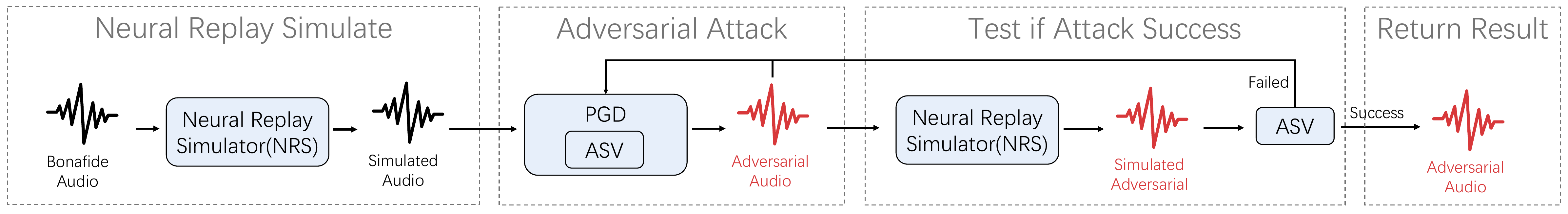}
    \caption{Pipeline of the NRS-based PGD Attack. The process begins with the Neural Replay Simulator (NRS) generating simulated audio from bona fide audio. This simulated audio is then used to create adversarial audio through the PGD algorithm targeting the ASV system. For ensemble attacks, multiple ASV models are targeted during the adversarial attack stage, and the adversarial audio must successfully fool all ASV models in the test stage. The adversarial audio is tested by simulating the OTA process again using the NRS to ensure the attack's effectiveness. If the attack is successful, the adversarial audio is returned as the final result; otherwise, the process iterates until a successful adversarial sample is generated.}
    \label{fig:nrs_pgd_pipeline}
\end{figure*}

\begin{algorithm}[htbp]
\caption{NRS-based OTA PGD Attack}
\label{alg:nrs_attack}
\begin{algorithmic}[1]
\REQUIRE Enrollment speech \(x_e\), evaluation speech \(x_v\), step size \(\alpha\), max steps \(S\), \(\epsilon\) for \(L_\infty\) norm ball, ASV model \(f(\cdot, \theta_{\text{ASV}})\) with threshold \(\tau\), pre-trained NRS model \(\tilde{o}(\cdot, \theta_{\text{NRS}})\)
\STATE $x_v^{replay} \gets \tilde{o}(x_v; \theta_{\text{NRS}})$ \COMMENT{Simulate OTA replay of clean input}
\STATE Initialize $x_1^{Adv} \gets x_v^{replay}$ \COMMENT{Initialize adversarial samples}
\FOR{$s=1$ \TO $S$}
\STATE $g \gets \nabla_{x_s^{Adv}} J(f(x_e, x_s^{Adv}; \theta_{\text{ASV}}))$ \COMMENT{Compute gradient w.r.t. model output}
\STATE $x_{s+1}^{Adv} \gets \text{clip}_{x_v, \epsilon}(x_s^{Adv} + \alpha \cdot \text{sign}(g))$ \COMMENT{PGD update with gradient}
\STATE $x_v^{Adv} \gets x_{s+1}^{Adv}$ \COMMENT{Update final adversarial sample after each iteration}
\STATE $x_{s+1}^{replay} \gets \tilde{o}(x_{s+1}^{Adv}; \theta_{\text{NRS}})$ \COMMENT{Simulate OTA replay within adversarial samples to test if attack is successful}
\IF{$f(x_e, x_{s+1}^{replay}; \theta_{\text{ASV}}) \geq \tau$}
\RETURN $x_v^{Adv}$ \COMMENT{Early stop if target speaker spoofed after OTA}
\ENDIF
\ENDFOR
\RETURN $x_v^{Adv}$ \COMMENT{Return final adversarial sample}
\end{algorithmic}
\end{algorithm}

To generate adversarial samples that can successfully attack the ASV system after over-the-air (OTA) transmission, we propose the NRS-based OTA Adversarial Attack method. The core idea is to integrate the adversarial sample generation process with OTA transmission simulation, ensuring that the generated adversarial samples can effectively fool the ASV model even in realistic OTA environments. Taking the PGD attack as an example, as illustrated in Algorithm~\ref{alg:nrs_attack} and Fig.~\ref{fig:nrs_pgd_pipeline}, the ensemble attack follows a similar principle and is not further elaborated.

Specifically, we first employ the pre-trained Neural Replay Simulator (NRS) to simulate OTA transmission on the original evaluation utterance $x_v$, introducing OTA transmission effects to obtain $x_v^{replay}$. We then use $x_v^{replay}$ as the initial adversarial input $x_0^{adv}$ for the PGD attack. At each iteration, we compute the gradient of the current adversarial sample $x_s^{adv}$ and perform the PGD update to obtain $x_{s+1}^{adv}$. To evaluate whether the generated adversarial sample $x_{s+1}^{adv}$ can successfully attack the ASV model after OTA transmission, we input it to the NRS model to simulate the OTA-transmitted speech $x_{s+1}^{replay}$. We then determine whether $x_{s+1}^{replay}$ can successfully fool the ASV model. If so, we terminate early and output $x_{s+1}^{adv}$ as the final adversarial sample $x_v^{adv}$.

By initiating the adversarial attack on the NRS-produced simulations, we ensure that the adversarial perturbations are specifically optimized for the conditions that the audio will encounter during OTA transmission. This method enhances the robustness and effectiveness of the adversarial samples, as they are crafted to not only exploit vulnerabilities in the target system but also to withstand the potential degradations introduced by the transmission process.

\section{CODA-OCC: Contrastive Domain-Aligned One-Class Classification for Adversarial Sample Detection}
\label{sec:adv_det}

Detecting adversarial samples poses a significant challenge due to the need for generalization in detection models. Adversarial perturbations are closely tied to the targeted model and attack algorithm, with continuous parameters leading to significant variations in their distribution. Even for the same targeted model, differences in training data and loss functions can result in substantial changes in the distribution of adversarial samples. As it is impossible to exhaustively enumerate all types of adversarial samples, detection models must possess high generalization capabilities to effectively handle the diverse and evolving nature of adversarial perturbations.

However, binary classification models often overfit to known adversarial samples. To address this issue, we leverage one-class classification (OCC)~\cite{ruff2018deep}, which trains solely on bona fide samples.

\subsection{Threat Model}
This work studies whitebox and blackbox attacks based on the adversary's knowledge of the ASV model. In the whitebox scenario, the adversary has full access to the model's internal details, while in the blackbox scenario, they do not. We focus on targeted attacks, aiming to mislead the ASV system into misclassifying a non-target speaker's utterance as a specific target speaker identity. For whitebox attacks, we use PGD and Ensemble PGD methods to generate adversarial samples, iteratively causing misclassification. For blackbox attacks, we adopt a transfer attack strategy, generating adversarial samples on a whitebox surrogate model to attack the blackbox model. Additionally, for ASV systems that only accept analog audio input, we employ over-the-air (OTA) attacks, where over-the-line adversarial samples are replayed through playback devices.

\subsection{One-class Classification based Adversarial Detection}

\begin{figure}[ht]
    \centering
    \includegraphics[width=1\linewidth]{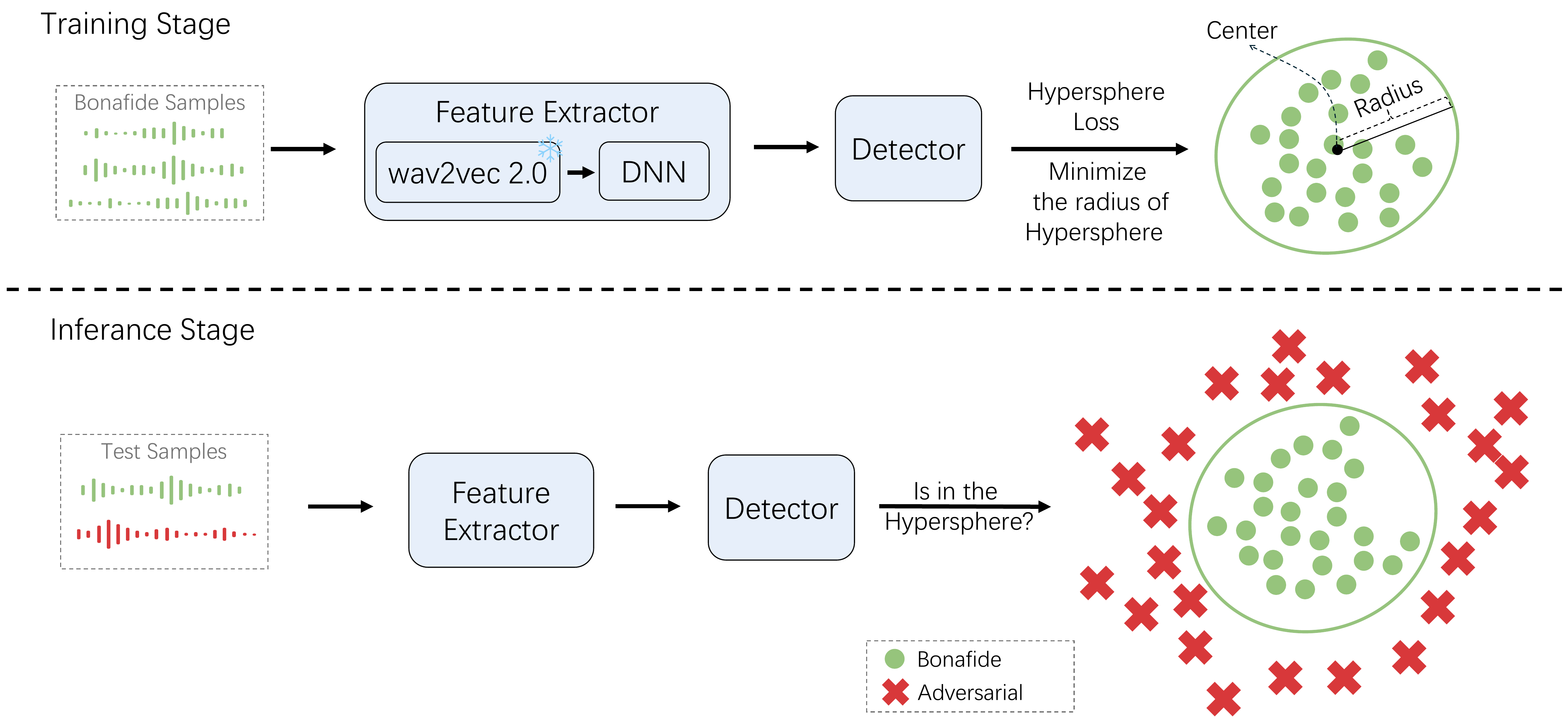}
    \caption{Architecture of the One-Class Classification-Based Adversarial Detection. During training, bona fide samples are used to learn the hypersphere parameters, ensuring minimal radius. During inference, the distance of test samples from the hypersphere center determines their classification as either bona fide or adversarial.}
    \label{fig:occ}
\end{figure}

This method focuses on modeling the distribution of bona fide samples and identifying any deviations from this distribution as potential adversarial attacks. \textbf{The key concept is to learn a hypersphere in high-dimensional feature space, positioning training samples close to the sphere's center while minimizing its radius.} By concentrating on the characteristics of bona fide data, one-class classification can achieve high generalization and effectively detect adversarial samples across different attack algorithms and model variations. As illustrated in Fig.~\ref{fig:occ}, the training phase determines the hypersphere's radius and center parameters. During inference, we calculate the distance from the test sample to the hypersphere center; if the distance exceeds the radius, the sample is classified as adversarial, otherwise, it is considered bona fide.

Formally, the objective can be expressed as:
\begin{equation}
\label{loss:hypersphere}
\min_{R, c} \, R^2 + \frac{1}{n} \sum_{i=1}^n \max \{0, \| f(f(x_i; \theta_{\text{Feat}}); \theta_{\text{Det}}) - c \|^2 - R^2 \}
\end{equation}
where \( R \) is the radius of the hypersphere, \( c \) is the center of the hypersphere, \( f(f(x_i; \theta_{\text{Feat}}); \theta_{\text{Det}}) \) is the output of the detector for the \(i\)-th bona fide sample, and \( n \) is the number of training samples.

During inference, for a test sample \( x \), its distance \( d(x) \) from the hypersphere center \( c \) is calculated as:
\begin{equation}
d(x) = \| f(f(x; \theta_{\text{Feat}}); \theta_{\text{Det}}) - c \|
\end{equation}
The sample \( x \) is classified as adversarial if \( d(x) > R \); otherwise, it is considered bona fide.

\subsection{Contrastive One-Class Classification (CO-OCC)}

\begin{figure*}[ht]
    \centering
    \includegraphics[width=0.8\linewidth]{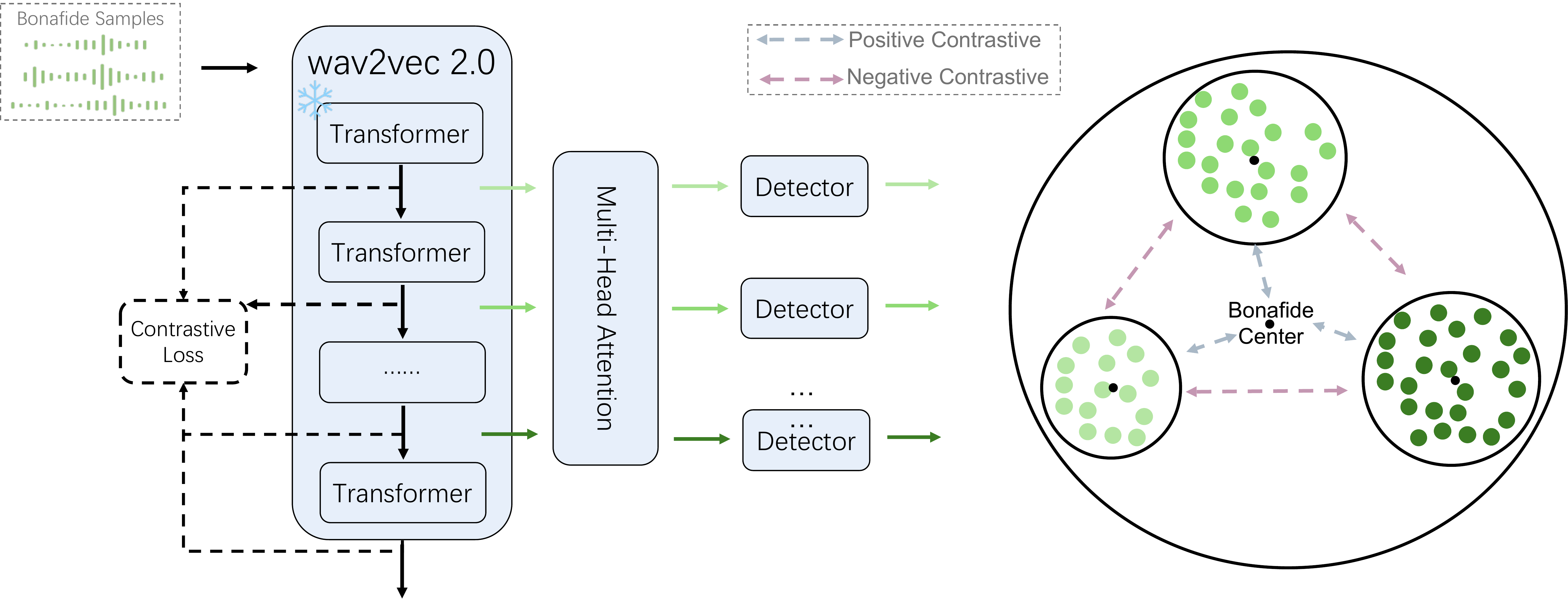}
    \caption{Architecture of the proposed Contrastive One-Class Classification (CO-OCC) method. The wav2vec 2.0 model extracts features from multiple layers, which are then used for contrastive learning. The core idea is to maintain each layer's features close to its hypersphere center and the bona fide center, while keeping different layers' centers far apart.}
    \label{fig:coocc}
\end{figure*}

Previous research showed that pre-trained models, such as wav2vec 2.0~\cite{baevski2020wav2vec}, contain different information across layers. Specifically, the earlier layers in wav2vec 2.0 models encode acoustic information, the next set of layers encodes phonetic class information, followed by word meaning information, before reverting back to encoding phonetic/acoustic information~\cite{mohamed2022self}. Therefore, features from different layers contain distinct characteristics relevant to adversarial sample detection.

This paper proposes the Contrastive One-Class Classification (CO-OCC) method, as shown in Fig.~\ref{fig:coocc}. \textbf{The core idea is to train a separate hypersphere for the features of each layer, ensuring that each layer's features are close to their respective hypersphere centers and the bona fide center, while keeping the centers of different layers' hyperspheres far apart. This approach preserves the unique representation of each layer while maintaining the decision-making capability of one-class classification.}

\subsection{Domain-Aligned One-Class Classification (DA-OCC)}

Traditional one-class classification (OCC) methods are insufficient for addressing the issue of adversarial attack detection due to intrinsic variations among bona fide samples. These variations arise from factors such as recording environment, compression encoding methods, and other external influences. For instance, the Libri-Light~\cite{Kahn_2020} dataset, which consists of audiobooks, features stable reading with minimal noise, whereas the VoxCeleb2~\cite{Chung_2018} dataset, sourced from YouTube, includes background noise and multiple speakers. Despite both datasets containing bona fide samples, their distributions differ significantly, which we term as \textit{Bona fide Intrinsic Variations}.

\begin{figure*}[t]
    \centering
    \subfloat[One-class domain alignment. Hyperspheres are learned for each domain and then aligned to handle adversarial samples effectively.]{\includegraphics[width=0.35\textwidth]{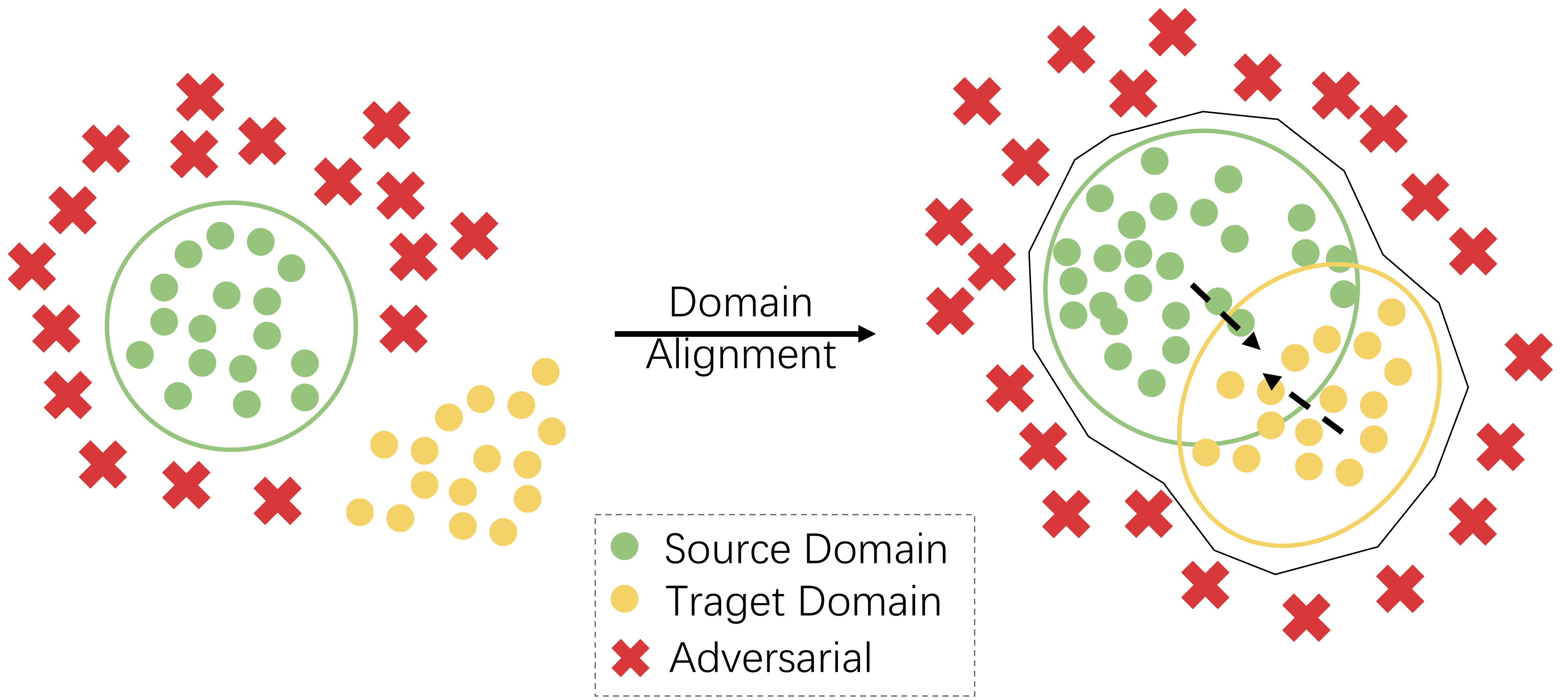}%
    \label{fig:daocc_1}}
    \hfil
    \subfloat[Architecture of the proposed Domain-Aligned One-Class Classification (DA-OCC). The feature extractor extracts features from bona fide samples of different domains, while hypersphere loss and MMD loss ensure alignment in decision and feature spaces.]{\includegraphics[width=0.55\textwidth]{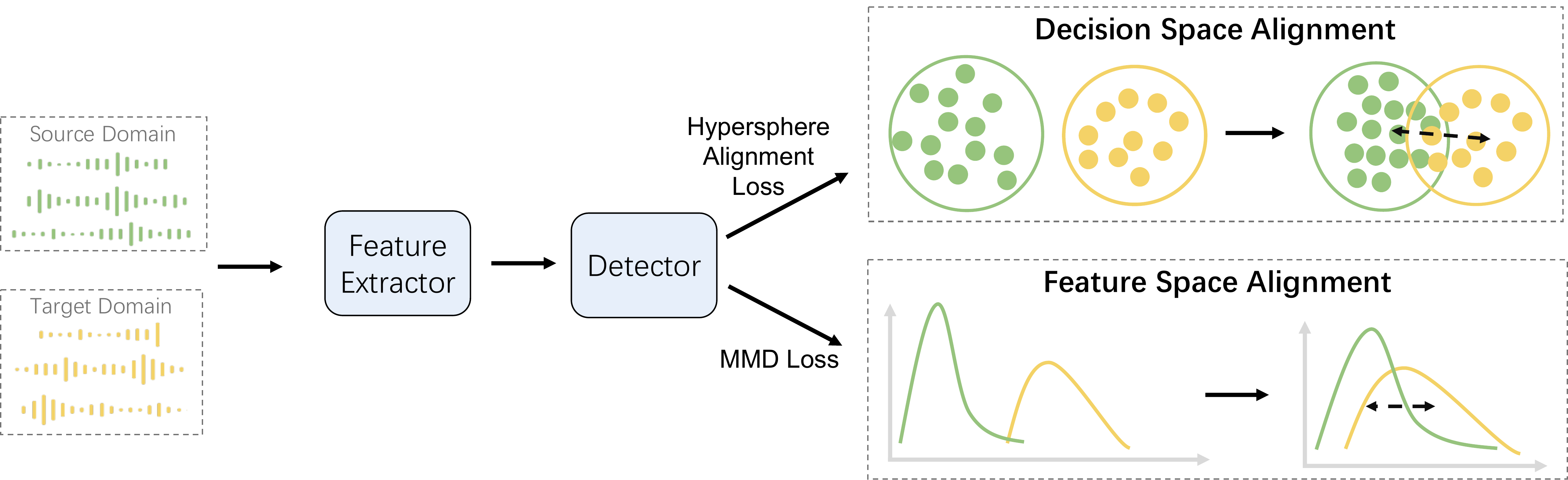}%
    \label{fig:daocc_2}}
    \caption{Illustrations of (a) one-class domain alignment and (b) the architecture of the proposed Domain-Aligned One-Class Classification (DA-OCC) model.}
    \label{fig:daocc}
\end{figure*}

To address this challenge, we propose the Domain-Aligned One-Class Classification (DA-OCC) method, as shown in Fig.~\ref{fig:daocc}. The primary goal of DA-OCC is to achieve high generalization of bona fide samples through domain alignment, as illustrated in Fig.~\ref{fig:daocc_1}, thereby enhancing the model's ability to detect adversarial samples across diverse domains.

Domain alignment is achieved by aligning both the decision space and the feature space, utilizing hypersphere alignment loss and MMD loss~\cite{gretton2006kernel}, as illustrated in Fig.~\ref{fig:daocc_2}. \textbf{The core idea is to constrain the centers of the hyperspheres for the source and target domains to be close, and to ensure that the feature distributions between domains are similar.} Additionally, as with the objective in Equation~\ref{loss:hypersphere}, the radii of the hyperspheres for both the source and target domains should be small, and the features within each domain should be close to their respective hypersphere centers.

Formally, the objective can be expressed as:
\begin{footnotesize}
\begin{equation}
\begin{aligned}
\min_{R_S, R_T, c_S, c_T} \, & R_S^2 + R_T^2 \\
& + \frac{1}{n_S} \sum_{i=1}^{n_S} \max \{0, \| f(f(x_i^S; \theta_{\text{Feat}}); \theta_{\text{Det}}) - c_S \|^2 - R_S^2 \} \\
& + \frac{1}{n_T} \sum_{j=1}^{n_T} \max \{0, \| f(f(x_j^T; \theta_{\text{Feat}}); \theta_{\text{Det}}) - c_T \|^2 - R_T^2 \} \\
& +  \| c_S - c_T \|^2 \\
& +  \text{MMD}(f(x^S; \theta_{\text{Feat}}), f(x^T; \theta_{\text{Feat}}))
\end{aligned}
\end{equation}
\end{footnotesize}
where \( R_S \) and \( R_T \) are the radii of the hyperspheres for the source and target domains, respectively, \( c_S \) and \( c_T \) are the centers of the hyperspheres for the source and target domains, respectively, \( f(f(x_i^S; \theta_{\text{Feat}}); \theta_{\text{Det}}) \) is the output of the detector for the \(i\)-th bona fide sample from the source domain, \( f(f(x_j^T; \theta_{\text{Feat}}); \theta_{\text{Det}}) \) is the output of the detector for the \(j\)-th bona fide sample from the target domain, \( n_S \) and \( n_T \) are the number of training samples in the source and target domains, and \( \text{MMD} \) represents the Maximum Mean Discrepancy between the feature distributions of the source and target domains.

\subsection{Contrastive Domain-Aligned One-Class Classification (CODA-OCC)}
Integrating the strengths of both Contrastive One-Class Classification (CO-OCC) and Domain-Aligned One-Class Classification (DA-OCC), this paper presents the Contrastive Domain-Aligned One-Class Classification (CODA-OCC) method. This novel approach is designed to improve adversarial sample detection across various domains. It achieves this by simultaneously preserving the unique information inherent in each layer of the feature extraction process, a benefit derived from contrastive learning, and by enhancing the model's overall generalization capability through robust domain alignment. The CODA-OCC method thus aims to provide a more robust and adaptable solution for detecting adversarial samples in complex, real-world scenarios where diverse data distributions are common.

\section{Adversarial Attack Experiments}
\subsection{Over-the-line Adversarial Attack Setups}
\subsubsection{X-Vector Speaker Verification}
\label{x-vector}
The current state-of-the-art ASV models are based on deep neural networks, where the speaker embeddings extracted by DNNs are generally referred to as x-vectors. In this work, we select four representative ASV models: ECAPATDNN~\cite{desplanques2020ecapatdnn}, XVector~\cite{snyder2018xvectors}, ResNetSE34V2~\cite{kwon2021ins}, and RawNet3~\cite{jung2022pushing}. In this paper, these four models are referred to as ECAPA, XVector, ResNet, and RawNet, respectively. Among them, XVector was the first DNN-based speaker verification model. ECAPA and ResNet are two prominent convolutional neural network-based speaker verification models. The first three models utilize hand-crafted MFCC features as audio representations, while RawNet learns to extract features directly from raw waveforms using a deep neural network.

To align with existing models, we adopt the common practice of training the ASV systems on VoxCeleb2~\cite{Chung_2018} and testing them on both the Voxceleb1~\cite{nagrani2017voxceleb} verification test set and Libri-Light. The results, as shown in Table~\ref{tab:asv_results}, demonstrate that the ASV systems perform as expected, confirming that the models are effective. Testing ASV performance ensures that the attacks target a valid model. Note that Libri-Light does not have an official ASV test set; the construction method is detailed in Section~\ref{dataset}.


\begin{table}[htbp]
\centering
\caption{Equal Error Rates (EERs,\%) of different speaker verification models on Voxceleb1 and Libri-Light datasets.}
\label{tab:asv_results}
\begin{tabular}{cccc}
\toprule
Model & EER (Voxceleb1) & EER (Libri-Light) \\
\midrule
ECAPA  & 1.26 & 1.15 \\
RawNet  & 1.06 & 0.95 \\
XVec  & 1.03 & 1.45 \\
ResNet & 2.08 & 1.01 \\
\bottomrule
\end{tabular}
\end{table}

\subsubsection{Dataset for Generating Adversarial Samples}
\label{dataset}
To generate adversarial samples, we utilized the Libri-Light (large) corpus. This corpus was chosen due to its large number of speakers (approximately 7,400) and its compliance with legal requirements. However, the Libri-Light (large) corpus contains an excessive number of samples, including some long sentences, which do not align with the specifications of ASV systems and make the adversarial attack more challenging. Consequently, we performed downsampling and simple filtering on the corpus. Specifically:
\begin{enumerate}
\item For every speaker, we randomly selected two samples approximately 4 seconds in duration.
\begin{enumerate}
\item Samples with less than two segments were discarded.
\item Samples containing only long segments were discarded.
\end{enumerate}
\item The two segments were designated as ``same''($y_{spk}=1$) and another speaker's utterance was randomly sampled as ``different''($y_{spk}=0$).
\end{enumerate}

Based on the aforementioned downsampling method, we ultimately retained 5,669 speakers, resulting in the construction of 11,338 speaker verification sample pairs. 

\subsubsection{Projected Gradient Descent (PGD) Attack}
The PGD attack is configured with a step size ($\alpha$) of 0.0005 and uses cosine similarity as the loss function. The maximum perturbation allowed is $\epsilon = 0.01$. The attack is performed for a maximum of 20 iterations ($S$), where the perturbation is clipped to ensure it stays below $\epsilon$ after each iteration.

\subsubsection{Ensemble PGD Attack}
For the \textit{Ensemble PGD Attack}, \textbf{three ASV models are used as victim models to generate adversarial samples, while the remaining one serves as a test for transfer attacks.} The PGD attack settings are the same as described previously for all victim models.

\subsection{Over-the-air Adversarial Attack Setups}
\subsubsection{Acoustic Environment and Equipment}
An Over-the-air(OTA) adversarial attack involves a perturbation generation algorithm, a loudspeaker, a microphone, and a replaying environment. In this work, we simulated the OTA adversarial attack in a soundproof studio to reduce the impact of environmental noise and focus the dataset on the impact of perturbation generation, loudspeakers, and microphones. These three variables already result in a significant number of combinations.

We chose three types of loudspeakers and three types of recording devices (i.e., microphones). The high-end, medium-end, and low-end loudspeakers are priced at around \$300 USD, \$90 USD, and \$50 USD, respectively. For the recording devices, we chose mobile devices common in daily lives. The iOS, Android-high, and Android-low devices are priced at around \$900 USD, \$750 USD, and \$310 USD, respectively.

The distance and angle between the microphone and loudspeaker are other factors. In this study, we simplified this factor. The distance between the loudspeaker and microphone is set to 0.3 meters, and the angle is set to 90 degrees.

\subsubsection{Neural Replay Simulator}
We used the VCTK dataset~\cite{veaux2016vctk} for training and evaluation. The dataset consists of speech recordings from 109 English speakers with various accents. After preprocessing, we selected 103 speakers with a total of 10,300 utterances (approximately 12.5 hours of speech). The preprocessing steps included:

\begin{itemize}
    \item Selecting the Sennheiser MKH 800 microphone recordings, as the DPA 4035 omni-directional microphone had low-frequency noise issues. The MKH 800 is a small diaphragm condenser microphone with a wide bandwidth.
    \item Excluding speakers p280 and p315 due to their MKH 800 audio recordings, leaving 103 speakers.
    \item Limiting the number of utterances to 100 per speaker due to time constraints.
\end{itemize}
The final dataset was split into 9,000 utterances for training and 1,300 utterances for testing.
The specific STFT configurations used in our experiments were FFT sizes of [128, 256, 512, 1024, 2048], hop sizes of [32, 64, 128, 256, 512], and window lengths of [128, 256, 512, 1024, 2048].
Conducting NRS-based OTA experiments for all ASV models is a massive undertaking. We believe that audio features significantly impact the performance of adversarial attacks. Therefore, we selected the two ASV models, RawNet and ECAPA, as discussed in Section~\ref{x-vector}. RawNet directly learns features from raw waveforms, while ECAPA employs hand-crafted MFCC features. Specifically, for the NRS-based OTA PGD attack, we conducted adversarial attacks on RawNet and ECAPA. For the NRS-based OTA Ensemble PGD attack, we consider two scenarios: without RawNet and without ECAPA.

\subsection{Statistics of AdvSV 2.0 Dataset}
Table~\ref{tab:advsv_stat} presents the statistics of the AdvSV 2.0 dataset, categorized by attack method, model, playback device, and record device. The \textbf{total number of samples is 629,735}, with a \textbf{total duration of 799.5 hours}. This breakdown provides insights into the dataset's composition across various categories, highlighting the distribution of samples and their respective durations.

\begin{table}[ht]
\centering
\caption{Statistics of the AdvSV 2.0 dataset, categorized by attack method, model, playback device, and record device. The table shows the number of samples and the total duration in hours for each category.}
\begin{tabular}{llll}
\toprule
\textbf{Category} & \textbf{Type} & \textbf{Samples} & \textbf{Hours} \\
\midrule
TOTAL & - & 629,735 & 799.5 \\
\midrule
\multirow{4}{*}{ATTACKMETHOD} 
 & PGD & 226,760 & 288.0 \\
 & Ensemble PGD & 226,760 & 288.0 \\
 & NRS PGD & 98,270 & 124.7 \\
 & NRS Ensemble PGD & 77,945 & 98.8 \\
\midrule
\multirow{8}{*}{MODEL} 
 & XVec & 56,690 & 72.0 \\
 & ResNet & 56,690 & 72.0 \\
 & RawNet & 103,954 & 132.0 \\
 & ECAPA & 107,696 & 136.8 \\
 & w/o XVec & 56,690 & 72.0 \\
  & w/o ResNet & 56,690 & 72.0 \\
 & w/o RawNet & 103,170 & 130.9 \\
  & w/o ECAPA & 88,155 & 111.9 \\
\midrule
\multirow{4}{*}{PLAYBACKDEVICE} 
& NA & 45,352 & 57.5 \\
 & High & 194,932 & 247.5 \\
  & Medium & 198,504 & 252.1 \\
 & Low & 190,947 & 242.4 \\
\midrule
\multirow{4}{*}{RECORDDEVICE} 
& NA & 45,352 & 57.5\\
 & iOS & 200,938 & 255.3 \\
 & AndroidHigh & 195,094 & 247.8 \\
 & AndroidLow & 188,351 & 239.0 \\
\bottomrule
\end{tabular}
\label{tab:advsv_stat}
\end{table}

\subsection{Over-the-line Adversarial Attack Results}

\begin{table}[ht]
\centering
\caption{Adversarial Attack Success Rates (\%)}
\begin{tabular}{cccccc}
\toprule
\multirow{2}{*}{\makecell{Attack \\ Method}} & \multirow{2}{*}{\makecell{Surrogate \\ Model}} & \multicolumn{4}{c}{Victim Model} \\ \cmidrule(l){3-6} 
                                &                            & RawNet & ECAPA & ResNet & XVec \\ \midrule
\multirow{4}{*}{PGD}           & RawNet                     & 100    & 14.3  & 11.2   & 23   \\
                                & ECAPA                      & 72     & 100   & 49.1   & 78.2 \\
                                & ResNet                     & 36.9   & 41.8  & 100    & 62.4 \\
                                & XVec                       & 51.1   & 56.7  & 45.1   & 100  \\ \midrule
\multirow{4}{*}{\makecell{Ensemble \\ PGD}}  & w/o RawNet                 & 88.9   & 100   & 100    & 100  \\
                                & w/o ECAPA                  & 100    & 70.3  & 100    & 100  \\
                                & w/o ResNet                 & 100    & 100   & 66.7   & 100  \\
                                & w/o XVec                   & 100    & 100   & 100    & 88.2 \\ \bottomrule
\end{tabular}
\label{tab:adv attack}
\end{table}

Table~\ref{tab:adv attack} details the success rates of adversarial attacks using different surrogate models against various victim models, employing distinct attack methodologies (PGD and Ensemble PGD). Each combination of surrogate and victim model, as well as the attack method used, represents a unique scenario in which the effectiveness of the adversarial attack is evaluated. The results across these scenarios provide insights into the robustness of different models under adversarial conditions. 

\textbf{White-box Attacks:} In scenarios where the surrogate and victim models are the same (e.g., RawNet as both surrogate and victim), the attack is considered a white-box attack, achieving a success rate of 100\%. This high success rate indicates the vulnerability of models to attacks where the adversary has complete knowledge of the model architecture and parameters. Notably, for Ensemble PGD attacks, all non-diagonal elements also represent white-box scenarios where the success rate reaches 100

\textbf{Transfer Attacks:} These attacks involve using a surrogate model to generate adversarial samples that are then used to attack a different victim model. For instance, adversarial samples created with ECAPA as the surrogate achieve a 72\% success rate when attacking RawNet. These results illustrate the variability in success rates among different model architectures, indicating different levels of transferability.

\textbf{Enhanced Transferability with Ensemble PGD:} By employing an ensemble of models (excluding the victim model) as surrogates, the success rate of attacking RawNet improved to 88.9\%. This significant enhancement in the transferability of adversarial samples stands in stark contrast to the highest success rate of 72\% achieved using a single surrogate model (ECAPA) in a standard PGD transfer attack. This trend of improved effectiveness with ensemble approaches is consistent across other victim models as well. For instance, when other models are targeted, the success rates with ensemble attacks generally reach or exceed those achieved with single model surrogates, demonstrating the robustness and efficiency of ensemble strategies in adversarial settings.

\subsection{Over-the-air Adversarial Attack Results}
\label{res:ota adv}

In our study, we comprehensively assessed the efficacy of adversarial attacks when subjected to over-the-air (OTA) transmission, a scenario that introduces real-world physical conditions such as air attenuation, device distortions, and environmental noise to the adversarial samples. Table~\ref{tab:adv_ota_diff} presents the average reduction in attack success rates of adversarial attacks post-OTA transmission, where results from 9 different device combinations are aggregated to facilitate a clearer understanding.

\begin{table}[ht]
\centering
\caption{Average Drop in Adversarial Attack Success Rates (\%) after OTA}
\label{tab:adv_ota_diff}
\begin{tabular}{cccccc}
\toprule
\multirow{2}{*}{\makecell{Attack \\ Method}} & \multirow{2}{*}{\makecell{Surrogate \\ Model}} & \multicolumn{4}{c}{Victim} \\
\cmidrule(l){3-6} 
 &  & RawNet & ECAPA & ResNet & XVec \\
\midrule
\multirow{4}{*}{\makecell{OTA \\ PGD}} & RawNet & -66.5 & -3.5 & -2.2 & -6.2 \\
 & ECAPA & -12.6 & 0.0 & -4.0 & -12.5 \\
 & ResNet & -8.0 & -7.5 & 0.0 & -12.4 \\
 & XVec & -7.8 & -9.2 & -3.2 & 0.0 \\
\midrule
\multirow{4}{*}{\makecell{OTA \\ Ensemble \\ PGD}} & w/o RawNet & -9.4 & 0.0 & -1.3 & -0.4 \\
 & w/o ECAPA & -19.9 & -12.5 & -0.4 & -0.5 \\
 & w/o ResNet & -10.7 & -0.3 & -6.4 & -0.5 \\
 & w/o XVec & -13.1 & -0.2 & -6.3 & -12.4 \\
\bottomrule
\end{tabular}
\end{table}

\textbf{Widespread decline in performance for OTA-transmitted adversarial samples:} Adversarial samples consistently show a significant decline in performance after undergoing OTA transmission. This observation highlights the considerable impact of physical distortions, such as air attenuation and device-induced noise, that adversarial samples encounter during real-world propagation.

\textbf{Relative robustness of white-box attacks:} Despite the overall decrease in the effectiveness of adversarial attacks with OTA transmission, white-box attacks demonstrate a relatively higher resilience, maintaining success rates that suggest a degree of robustness against the physical distortions imposed by the OTA environment.

\textbf{Significant susceptibility of RawNet to OTA distortions, influenced by its front-end design:} Adversarial samples crafted using RawNet as the surrogate model exhibit pronounced susceptibility to the degradations caused by OTA transmission. This increased vulnerability is largely attributed to RawNet's reliance on a learnable front-end, which, unlike the Mel-frequency cepstral coefficients (MFCC) used by other models, is less robust to the physical distortions typically encountered in real-world scenarios. MFCCs, being more closely aligned with human auditory perceptions, offer enhanced robustness against such distortions, thereby providing better performance stability.

\subsection{NRS based OTA Adversarial Attack Results}

\textbf{NRS-based attack methods improve the performance of adversarial attack in both black-box and white-box scenarios.} In Section~\ref{res:ota adv}, we observe the significant susceptibility of RawNet to OTA distortions. The results of NRS-based attacks are shown in Table~\ref{tab:nrs adv ota}. After applying NRS, the attack success rate of adversarial samples increased from 33.5\% to 66.9\%, which is a substantial improvement. Additionally, for ECAPA, the attack success rate remained at 100\% regardless of whether NRS was applied, indicating that NRS does not compromise the effectiveness of adversarial samples. In terms of transfer attacks, both w/o RawNet and w/o ECAPA showed performance improvements, with w/o ECAPA achieving an absolute performance increase of 18\%.

\begin{table*}[ht]
\centering
\caption{Impact of Neural Replay Simulator (NRS) on Over-the-air (OTA) Adversarial Attack Success Rates (\%)}
\label{tab:nrs adv ota}
\begin{tabular}{cccccccc}
\toprule
\multirow{2}{*}{\textbf{Attack Method}} & \multirow{2}{*}{\textbf{Surrogate Model}} & \multirow{2}{*}{\textbf{Victim Model}} & \multirow{2}{*}{\textbf{Playback Device}} & \multicolumn{3}{c}{\textbf{Record Device}} & \multirow{2}{*}{\textbf{Average Result}} \\
\cmidrule(lr){5-7}
 &  &  &  & \textbf{iOS} & \textbf{Android High} & \textbf{Android Low} &  \\
\midrule
\multirow{6}{*}{\makecell{OTA \\ Adversarial\\\textbf{w/o NRS}}} & \multirow{3}{*}{RawNet} & \multirow{3}{*}{RawNet} & High & 39.2 & 21.6 & 31.0 & \multirow{3}{*}{33.5} \\
 &  &  & Medium & 48.3 & 28.6 & 38.9 &  \\
 &  &  & Low & 34.1 & 27.5 & 32.5 &  \\
\cmidrule(lr){2-8}
 & \multirow{3}{*}{w/o RawNet} & \multirow{3}{*}{RawNet} & High & 84.3 & 70.9 & 82.1 & \multirow{3}{*}{79.5} \\
 &  &  & Medium & 84.7 & 80.2 & 82.2 &  \\
 &  &  & Low & 77.5 & 73.9 & 79.6 &  \\
\midrule
\multirow{6}{*}{\makecell{OTA \\ Adversarial\\\textbf{w/ NRS}}} & \multirow{3}{*}{RawNet} & \multirow{3}{*}{RawNet} & High & 61.6 & 40.4 & 64.0 & \multirow{3}{*}{66.9} \\
 &  &  & Medium & 83.8 & 75.6 & 76.9 &  \\
 &  &  & Low & 77.0 & 42.7 & 80.0 &  \\
\cmidrule(lr){2-8}
 & \multirow{3}{*}{w/o RawNet} & \multirow{3}{*}{RawNet} & High & 91.6 & 72.1 & 91.3 & \multirow{3}{*}{80.8} \\
 &  &  & Medium & 90.3 & 74.2 & 84.1 &  \\
 &  &  & Low & 82.6 & 66.0 & 74.9 &  \\
\midrule
\multirow{6}{*}{\makecell{OTA \\ Adversarial\\\textbf{w/o NRS}}} & \multirow{3}{*}{ECAPA} & \multirow{3}{*}{ECAPA} & High & 100.0 & 100.0 & 100.0 & \multirow{3}{*}{100.0} \\
 &  &  & Medium & 100.0 & 100.0 & 100.0 &  \\
 &  &  & Low & 100.0 & 100.0 & 99.9 &  \\
\cmidrule(lr){2-8}
 & \multirow{3}{*}{w/o ECAPA} & \multirow{3}{*}{ECAPA} & High & 56.6 & 55.1 & 56.9 & \multirow{3}{*}{57.8} \\
 &  &  & Medium & 59.9 & 60.6 & 59.9 &  \\
 &  &  & Low & 57.4 & 59.0 & 54.9 &  \\
\midrule
\multirow{6}{*}{\makecell{OTA \\ Adversarial\\\textbf{w/ NRS}}} & \multirow{3}{*}{ECAPA} & \multirow{3}{*}{ECAPA} & High & 100.0 & 100.0 & 100.0 & \multirow{3}{*}{100.0} \\
 &  &  & Medium & 100.0 & 100.0 & 100.0 &  \\
 &  &  & Low & 100.0 & 100.0 & 100.0 &  \\
\cmidrule(lr){2-8}
 & \multirow{3}{*}{w/o ECAPA} & \multirow{3}{*}{ECAPA} & High & 74.8 & 68.7 & 86.9 & \multirow{3}{*}{75.8} \\
 &  &  & Medium & 74.3 & 71.3 & 77.7 &  \\
 &  &  & Low & 71.1 & 74.3 & 83.3 &  \\
\bottomrule
\end{tabular}
\end{table*}

\section{Adversarial Sample Detection Experiments}

\subsection{Setups}

\subsubsection{Baseline Method}
The baseline method follows~\cite{wu2022adversarial}\footnote{\href{https://github.com/hbwu-ntu/spot-adv-by-vocoder}{https://github.com/hbwu-ntu/spot-adv-by-vocoder}}, utilizing the current SOTA architecture (as shown in Fig.~\ref{fig:adv_det}). It employs ParallelWaveGAN~\cite{yamamoto2020parallel} for adversarial purification, with the ASV structure based on ResNet.

\subsubsection{Dataset Split for Adversarial Sample Detection}
\begin{table}[ht]
\centering
\caption{Adversarial Detection Dataset}
\begin{tabular}{llrrr}
\hline
\textbf{} & \textbf{Dataset} & \textbf{Speakers} & \textbf{Utterances} & \textbf{Hours} \\
\hline
\multirow{2}{*}{\textbf{Train Set}} & Libri-Light Medium & 500 & 120,410 & 556 \\
                   & Voxceleb2          & 459 & 120,146 & 258 \\
\hline
\multirow{2}{*}{\textbf{Dev Set}}   & Libri-Light Medium & 50  & 9,484   & 39 \\
                   & Voxceleb2          & 37  & 10,393  & 22 \\
\hline
\multirow{3}{*}{\textbf{Test Set}}  & Libri-Light Medium & 100 & 61,945  & 327 \\
                   & Voxceleb2          & 239 & 61,872  & 133 \\
                   & AdvSV 2.0              & 5,669 & 547,264 & 695 \\
\hline
\end{tabular}
\end{table}

The dataset used for the proposed adversarial detection algorithm is shown in the table. Our bona fide samples are sourced from Libri-Light (Medium)~\cite{Kahn_2020} and VoxCeleb2~\cite{Chung_2018}, which have significant channel differences, such as the presence of noise, encoding methods, and recording conditions, leading to domain mismatch issues.

In terms of speakers, these datasets include a large number of speakers, which helps to mitigate bias that could arise from a smaller speaker pool. Additionally, since AdvSV 2.0 is constructed based on Libri-Light (Large), we have removed overlapping speakers between Libri-Light Medium and Libri-Light Large beforehand. Furthermore, there is no overlap of speakers across the training, validation, and test sets.

\subsubsection{Training and Evaluation Setup}
The experiments were conducted using V100 GPUs. The model was trained for 10 epochs with an initial learning rate of 1e-4, which was reduced by a factor of ten every 3 epochs. The Adam optimizer was used for training. The model with the lowest Equal Error Rate (EER) on the validation set was selected as the final model. \textbf{In this study, we treat Libri-Light as the source domain and VoxCeleb2 as the target domain.} We evaluated the adversarial detection model using EER, AUC, FAR, and FRR. The threshold for EER was determined from validation set.

\section{Adversarial Sample Detection Results}

\begin{figure*}[t]
    \centering
    \subfloat[OCC clustering visualization results. The training set from Libri-Light (green) is well clustered together, but there is significant overfitting, leading to unknown bona fide samples being misclassified as adversarial.]{\includegraphics[width=0.3\textwidth]{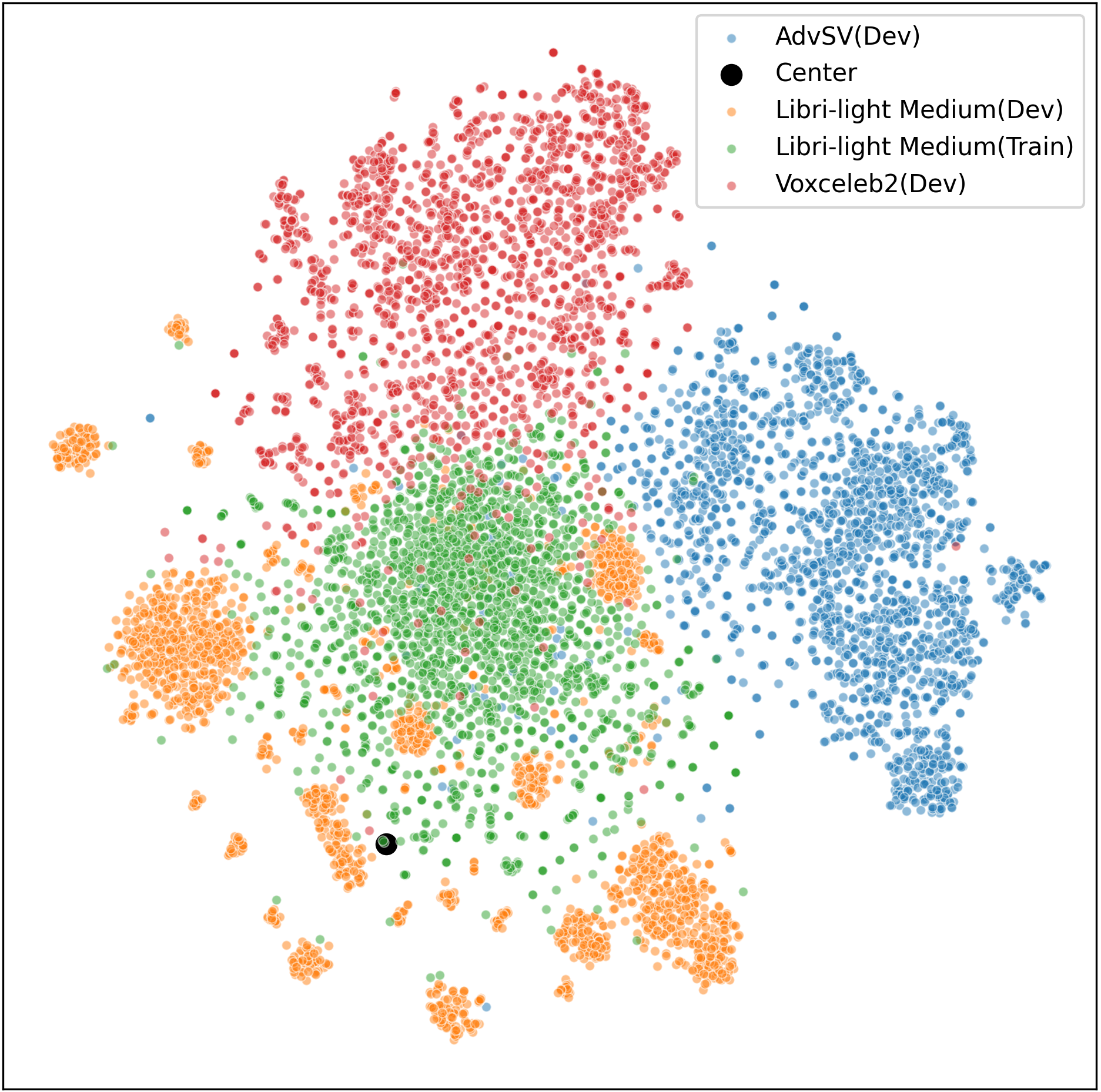}%
    \label{fig:visu_occ}}
    \hfil
    \subfloat[DA-OCC clustering visualization results. The source domain (Libri-Light) and target domain (VoxCeleb2) are well aligned, exhibiting a more consistent distribution in the feature space.]{\includegraphics[width=0.3\textwidth]{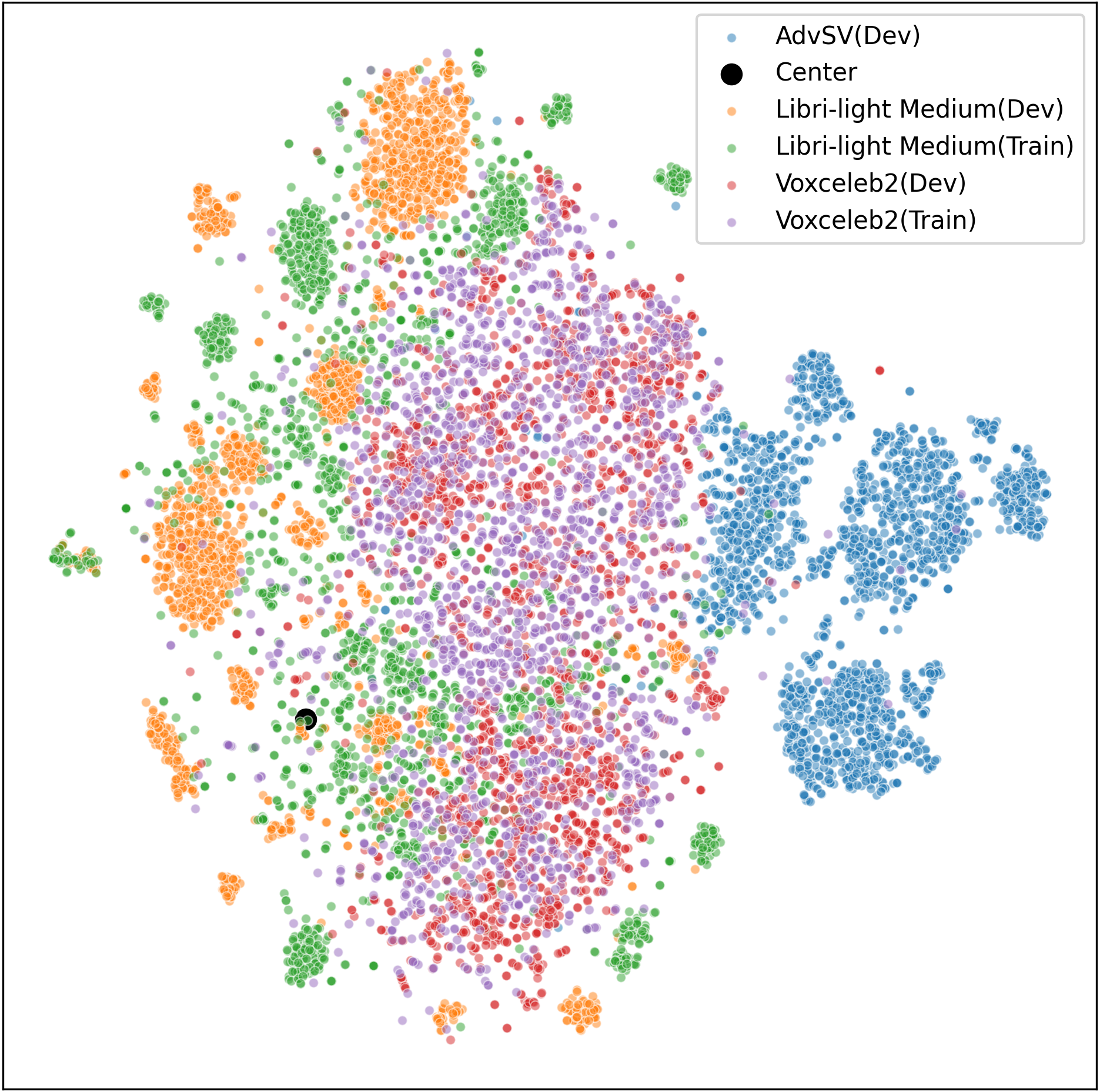}%
    \label{fig:visu_daocc}}
    \hfil
    \subfloat[CODA-OCC clustering visualization results. Compared to DA-OCC, it can be observed that the internal variations within the bona fide class are preserved.]{\includegraphics[width=0.3\textwidth]{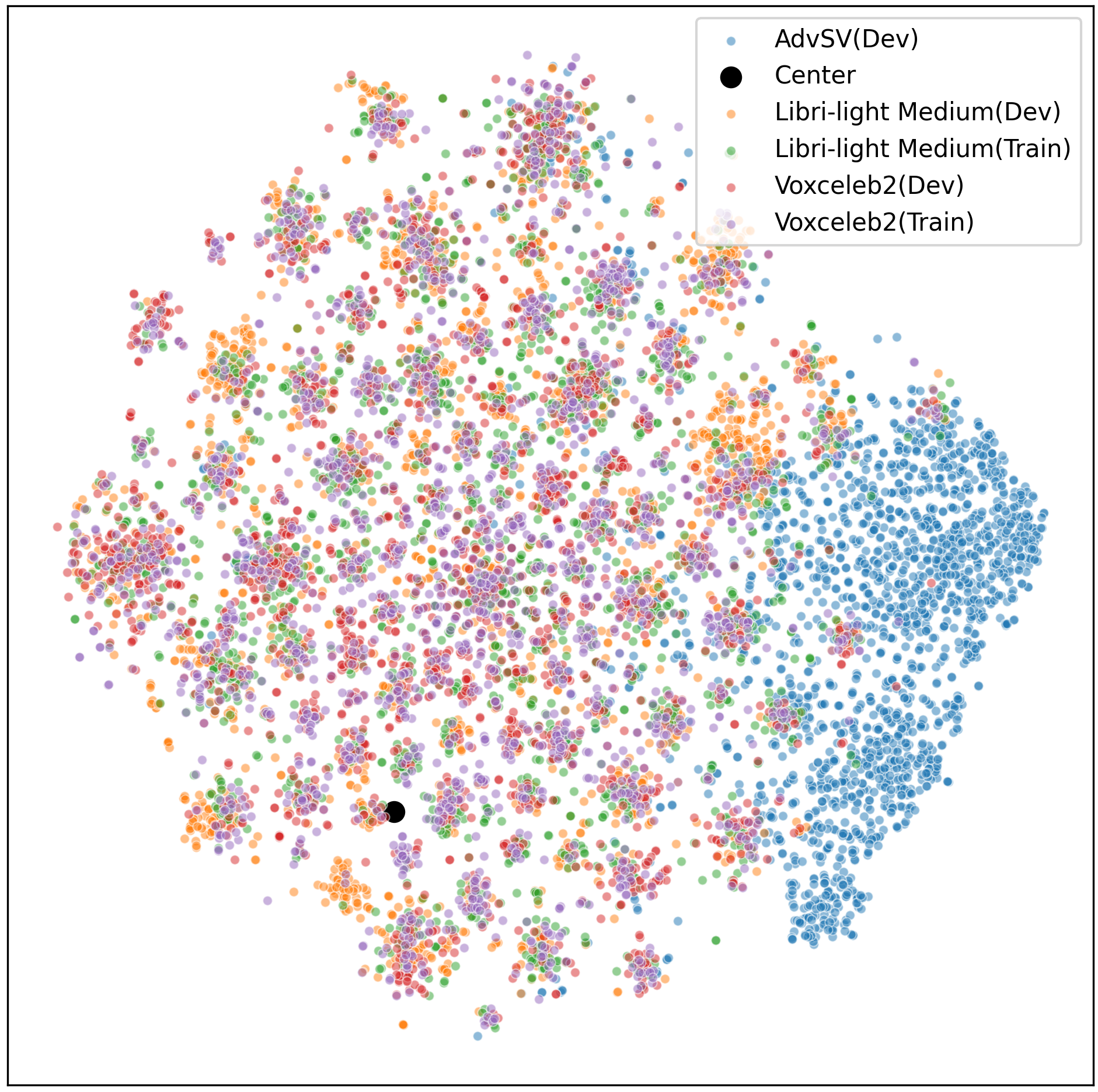}%
    \label{fig:visu_codaocc}}
    \caption{Clustering visualization results for different methods. In all plots, green represents Libri-Light Medium (Train), orange represents Libri-Light Medium (Dev), blue represents AdvSV 2.0, red represents VoxCeleb2 (Dev), and purple represents VoxCeleb2 (Train). VoxCeleb2 (Train) is used for alignment data.}
    \label{fig:visu}
\end{figure*}

Table~\ref{tab:adv_det_res} shows the adversarial detection results.
Fig.~\ref{fig:visu} shows the clustering visualization results for different methods by t-SNE~\cite{van2008visualizing}. Note that VoxCeleb2 is used for alignment data.

\begin{table*}[ht]
\centering
\caption{Performance of Adversarial Sample Detection (OCC: One-class Classification, CO: Contrastive Learning, DA: Domain Alignment)}
\label{tab:adv_det_res}
\begin{tabular}{llccccccc}
\toprule
\multirow{2}{*}{\textbf{Method}} & \multicolumn{2}{c}{\textbf{Dataset}} & \multirow{2}{*}{\textbf{EER(\%)}} & \multicolumn{3}{c}{\textbf{FAR(\%)}} & \multirow{2}{*}{\textbf{FRR(\%)}} & \multirow{2}{*}{\textbf{AUC}} \\
\cmidrule(lr){2-3} \cmidrule(lr){5-7}
 & \textbf{Train} & \textbf{Alignment} & & \textbf{ALL} & \textbf{Libri-Light} & \textbf{VoxCeleb2} & & \\
\midrule

Baseline~\cite{wu2022adversarial} & - & - & 19.8 & 12.1 & 5.2 & \textbf{17.3} & 30.0 & 0.90 \\
\midrule
OCC & Libri-Light & - & 37.4 & 36.8 & 8.0 & 65.7 & 38.1 & 0.67 \\

~~~~w/ CO & Libri-Light & - & 18.1 & 18.1 & 2.9 & 33.2 & 18.2 & 0.88 \\
~~~~w/ DA & Libri-Light & VoxCeleb2 & 13.5 & 13.0  & 5.0  & 20.9 & 14.0 & 0.93 \\
~~~~w/ CODA & Libri-Light & VoxCeleb2 & \textbf{11.2} & \textbf{11.3} & \textbf{2.6} & 19.9 & \textbf{11.2} & \textbf{0.95} \\
\bottomrule
\end{tabular}
\end{table*}

\textbf{The baseline method's white-box assumption leads to a higher FRR.} In the baseline method, it is assumed that the ASV model used during detection is the same as the one targeted by the adversarial attack. This assumption results in poor performance in transfer attacks, where adversarial samples generated for a different ASV model show less significant score changes. Consequently, these adversarial samples are not detected effectively, leading to a higher FRR.

\textbf{The OCC method performs well in in-domain tests but poorly in cross-domain tests.}
From Table~\ref{tab:adv_det_res}, it is evident that the OCC method performs well in in-domain tests (i.e., when the training and testing datasets are the same). For example, the model trained on Libri-Light achieves an FAR of only 8.0\% on Libri-Light, but a much higher FAR of 65.7\% on VoxCeleb2, indicating poor cross-domain performance. Fig.~\ref{fig:visu_occ} shows the OCC clustering visualization results. The training set from \textit{Libri-Light (train)} is well clustered together, but there is significant overfitting, leading to unknown bona fide samples being misclassified as adversarial, resulting in a high FAR.

\textbf{The CO-OCC method improves the model's generalization ability through contrastive learning.}
The CO-OCC method leverages the advantages of contrastive learning, enabling the model to better capture feature information at different levels, thereby enhancing its generalization ability. When trained solely on Libri-Light Medium, the CO-OCC method achieves an EER of 18.1\% and an AUC of 0.88, showing good performance, particularly with an FAR of only 2.9\% on Libri-Light.

\textbf{The domain alignment (DA-OCC) strategy significantly enhances the model's cross-domain performance.}
The DA-OCC method, which incorporates domain alignment during training, effectively reduces the feature distribution disparity between the source and target domains. This reduction allows the model to generalize better across different test sets. For instance, the DA-OCC method achieves an EER of 13.5\% and an AUC of 0.93. Fig.~\ref{fig:visu_daocc} shows the DA-OCC clustering visualization results, where the source domain (Libri-Light) and the target domain (VoxCeleb2) are well aligned, exhibiting a more consistent distribution in the feature space.

\textbf{Aligning either the decision space or the feature space can improve performance, but combining both results in the most significant performance enhancement.} As shown in Table~\ref{tab:adv_daocc_abl}, in the DA-OCC method, aligning only the decision space or the feature space both contribute to performance improvement. However, combining these two alignments leads to the highest performance gains, demonstrating the importance of addressing multiple aspects of domain alignment. Additionally, when jointly trained on Libri-Light and VoxCeleb2, the OCC method shows some improvement in overall performance (EER drops to 30.7\% and AUC increases to 0.77). However, this improvement is not as significant as that achieved by using domain alignment (DA). Joint training alone does not sufficiently address the feature distribution differences between domains, leading to unstable performance in cross-domain tests.

\begin{table}[ht]
\centering
\caption{Ablation Study of Domain Alignment (LL: Libri-Light, Vox2: VoxCeleb2)}
\label{tab:adv_daocc_abl}
\begin{tabular}{llccc}
\toprule
\multirow{2}{*}{\textbf{Method}} & \multicolumn{2}{c}{\textbf{Dataset}} & \multirow{2}{*}{\textbf{EER(\%)}} & \multirow{2}{*}{\textbf{AUC}} \\
\cmidrule(lr){2-3}
 & \textbf{Train} & \textbf{Alignment} & & \\
\midrule
OCC & LL+Vox2 & - & 30.7 & 0.77 \\
\midrule
DA-OCC & LL & Vox2 & \textbf{13.5} & \textbf{0.93} \\
~~~~w/o Align Decision &  LL & Vox2 & 16.7 & 0.91 \\
~~~~w/o Align Feature &  LL & Vox2 & 18.9 & 0.89 \\
\bottomrule
\end{tabular}
\end{table}

\textbf{The CODA-OCC method combines contrastive learning and domain alignment to achieve optimal adversarial sample detection performance.}
The CODA-OCC method combines the strengths of contrastive learning and domain alignment, resulting in significant performance improvement. It achieves an EER of 11.2\%, an AUC of 0.95, and an overall FAR of 11.3\%. The FARs on Libri-Light and VoxCeleb2 are 2.6\% and 19.9\%, respectively, and the FRR is reduced to 11.2\%. This significant performance improvement indicates that the CODA-OCC method excels in handling domain alignment and adversarial sample detection. Fig.~\ref{fig:visu_codaocc} shows the CODA-OCC clustering visualization results. Compared to DA-OCC, it can be observed that the internal variations within the bona fide class are preserved, enhancing the model's generalization ability.

\section{Conclusion}
In this work, we propose the AdvSV 2.0 dataset for evaluating adversarial attacks in speaker verification (ASV) systems. This dataset utilizes four mainstream ASV models to generate adversarial samples for attack.

To enhance the transferability of adversarial samples, we employed ensemble PGD adversarial attacks.
For transfer attacks, the success rate reached at least 66.7\%, with an average improvement of 14.5\% compared to non-ensemble attacks.
Adversarial attack performance consistently decreases after over-the-air (OTA) transmission. Therefore, we proposed a Neural Replay Simulator (NRS)-based adversarial attack method, which effectively enhances the attack performance of adversarial samples after OTA transmission. When using ECAPA as the victim model, the attack success rate increased by 18\% to 75.8\%. These experiments indicate that the AdvSV 2.0 dataset poses significant security threats to existing ASV systems, highlighting their vulnerability.

Exhaustively enumerating all adversarial samples is impractical due to the continuous nature of generation parameters. Consequently, binary classification models risk overfitting to known adversarial samples. We designed an adversarial sample detection method based on one-class classification. To address the intrinsic variability within bona fide samples, we employed transfer learning to align bona fide samples from different domains, reducing the EER by an absolute 23.9\%. Furthermore, we introduced a contrastive learning paradigm within the one-class classification framework, improving the EER by an additional 2.3\%, resulting in a final EER of 11.2\%.

In future work, we plan to incorporate more bona fide sample sets to enhance the robustness of one-class classification and further improve adversarial sample detection performance. Additionally, we aim to validate the effectiveness of CODA-OCC against other types of attacks.

\bibliographystyle{IEEEtran}
\bibliography{main}

\end{document}